\documentclass[11pt,draft]{article}
\usepackage{amssymb,amsmath}
\usepackage{graphicx}
\newcounter{proposition}

\newcommand{\nothing}[1]{}

\newcommand{\beq}[1]{\begin{equation}\label{#1}}
\newcommand{\eeq}{\end{equation}}
\newcommand{\req}[1]{(\ref{#1})}
\newcommand{\bmu}[1]{\begin{multline}\label{#1}}
\newcommand{\emu}{\end{multline}}

\newcommand{\eq}{\triangleq}

\renewcommand{\varlimsup}{\mathop{\overline{\lim}}\limits}

\newcommand{\x}{{\textbf{\textit{x}}}}

\renewcommand{\a}{{\textbf{\textit{a}}}}

\newcommand{\F}{\mathcal{F}}

\newcommand{\B}{{\bf B}}
\newcommand{\BB}{{\cal B}}
\newcommand{\A}{{\cal A}}
\newcommand{\T}{{\cal T}}

\renewcommand{\a}{{\bf a}}

\renewcommand{\u}{{\bf u}}
\renewcommand{\v}{{\bf v}}
\renewcommand{\S}{{\cal S}}
\renewcommand{\P}{{\cal P}}
\renewcommand{\L}{{\bf L}}
\newcommand{\E}{{\bf E}}

\renewcommand{\l}{\ell}

\renewcommand{\S}{{\mathcal{S}}}
\renewcommand{\L}{{\mathcal{L}}}

\newcommand{\s}{{ {s}}}

\renewcommand{\chi}{\upsilon}
\renewcommand{\l}{{ {L}}}
\newcommand{\0}{{\textbf{\textit{0}}}}

\renewcommand{\(}{\left(}
\renewcommand{\)}{\right)}
\renewcommand{\[}{\left[}
\renewcommand{\]}{\right]}

\renewcommand{\l}{\ell}

\begin{document}

\begin{center}
{\Large\bf  Almost Disjunctive List-Decoding  Codes}
 \\[15pt]
{\bf A. G. D'yachkov, \quad I.V. Vorobyev, \quad N.A. Polyanskii,\quad V.Yu. Shchukin}
\\[15pt]
Moscow State University, Faculty of Mechanics and
Mathematics,\\
Department of Probability Theory, Moscow, 119992, Russia,\\
{\sf agd-msu@yandex.ru,\quad vorobyev.i.v@yandex.ru,\quad nikitapolyansky@gmail.com,\quad
vpike@mail.ru}
\end{center}

\textbf{Abstract.}\quad
A binary code is said to be a disjunctive list-decoding $s_L$-code, $s\ge1$, $L\ge1$,
(briefly,  LD $s_L$-code) if  the  code is identified by  the incidence matrix of a
family of finite sets in which  the union of any $s$ sets can cover not more than $L-1$ other sets
of the family. In this paper, we introduce a natural  {\em probabilistic} generalization
of LD $s_L$-code  when the code is said to be an   almost disjunctive LD
$s_L$-code if  the unions of  {\em almost all} $s$ sets
satisfy the given condition. We develop a random coding method
based on the ensemble of binary constant-weight codes
to obtain  lower bounds on the capacity and error probability exponent of
such codes. For the considered ensemble our lower  bounds are asymptotically tight.

\medskip


\section{Notations and Definitions}
\quad
Let $N$, $t$, $s$, and $L$ be integers, where
 $1\le s<t$, $1\le L\le t-s$. Let
$\eq$ denote the equality by definition, $|A|$ -- the size of set $A$ and
$[N]\eq\{1,2,\dots,N\}$ - the set of integers from~$1$ to~$N$.
The standard symbol  $\lfloor a\rfloor$
($\lceil a\rceil$) will be used to denote the largest (least) integer $\le a$~($\ge a$).
A binary $(N\times t)$-matrix
\beq{X}
X=\|x_i(j)\|, \quad x_i(j)=0,1,\quad\x_i\eq(x_i(1),\dots,x_i(t)),\quad
\x(j)\eq(x_1(j),\dots,x_N(j)),
\eeq
 $i\in[N]$, $j\in[t]$,  with $N$ rows $\x_1,\dots,\x_N$ and $t$ columns
$\x(1),\dots,\x(t)$ (codewords) is called a {\em binary code of length $N$  and size $t=\lceil 2^{RN}\rceil$}
(briefly, {\em $(N,R)$-code}), where a fixed parameter $R>0$ is called the {\em rate} of code~$X$~\cite{f65}-\cite{ck85}.
For any code $X$ and  any subset $\S\subset[t]$ of size $|\S|=\s$, the symbol
$\x(\S)\eq\{\,\x(j)\,:\,j\in\S\}$ will denote the corresponding $s$-subset of codewords (columns)
of the code~$X$.  The number of $1$'s in column  $x(j)$, i.e.,
$|\x(j)|\eq\sum\limits_{i=1}^N\,x_i(j)$, is called
the {\em weight} of  $x(j)$,~$j\in[t]$.
A code $X$ is called a {\em constant weight} binary code of  weight $w$, $1<w<N$, if for any
$j\in[t]$, the weight~$|\x(j)|=w$.
The standard symbol $\bigvee$ denotes
the {\em disjunctive} (Boolean) sum of two binary numbers:
$$
0\bigvee0=0,\quad  0\bigvee 1=1\bigvee 0 = 1\bigvee 1 = 1,
$$
as well as the component-wise disjunctive  sum of two binary columns.
We say that a column
 $\u$   covers column $\v$ ($\u \succeq \v$) if~$\u\bigvee \v=\u$.

\textbf{Definition 1.}\quad An  $s$-subset of columns $\x(\S)$, $|\S|=\s$, of a code $X$
is said to be an {\em $s_L$-bad} subset of columns in the code $X$ if there exists a subset $\L\subset[t]$
of size $|\L|=L$, such that $\S\cap\L=\varnothing$ and the disjunctive sum
\beq{s_L}
\bigvee_{i\in\S}\x(i)\succeq \bigvee_{j\in\L}\x(j).
\eeq
Otherwise, the $s$-subset  $\x(\S)$ is called {\em $s_L$-good} subset of columns  in the code
$X$. In other words, for any $s_L$-good subset of columns  in a code $X$,
the disjunctive sum of its $s$ columns   can cover not more than  $L-1$ columns
of the code $X$ that are not components of the given $s$-subset.


\textbf{Definition 2.}\quad Let $\epsilon$, $0\le \epsilon<1$,
be a fixed parameter.
A code $X$ is said to be a  {\em disjunctive  list-decoding   $(s_L,\epsilon)$-code}
(or {\em almost disjunctive  list-decoding   $s_L$-code}) 
 of {\em strength}  $s$, {\em list size $L$} and {\em error probability} $\epsilon$, $0\le \epsilon<1$,
  (briefly, {\em LD  $(\;s_L,\,\epsilon)$-code}), if
the number ${\bf G}_L(s,X)$  of all $s_L$-good $s$-subsets of columns  of the code $X$ is at least~$(1-\epsilon)\cdot\,{t\choose s}$.
In other words, the number $\B_L(s,X)$ of all $s_L$-bad $s$-subsets of columns for  LD  $(\;s_L,\,\epsilon)$-code  $X$
does not exceed~$\epsilon\,{t\choose s}$, i.e.,
\beq{B_eps}
\B_L(s,X)\eq {t\choose s}-{\bf G}_L(s,X)\le \epsilon\cdot\,{t\choose s} \;
\Longleftrightarrow\;
\frac{\B_L(s,X)}{{t\choose s}}\,\le\,\epsilon
\eeq

The concept of LD $(\;s_L,\,\epsilon)$-code
can be considered as a natural generalization of the classical superimposed $s$-code
of Kautz-Singleton~\cite{ks64} corresponding to  the   case $L=1$ and~$\epsilon=0$.
For the  case $L\ge1$ and $\epsilon=0$,  disjunctive list-decoding codes (LD $s_L$-codes) were
investigated in works~\cite{dr82}-\cite{d03}
and the last detailed survey of the most important results obtained  for LD $s_L$-codes is
given in the recent paper~\cite{d14_pit} (see, also, preprint~\cite{d14}).

\textbf{Definition 3.}\quad
Let  $t_{\epsilon}(N,s,L)$ be the maximal size of LD $(\;s_L,\,\epsilon)$-codes of length $N$ and
let $N_{\epsilon}(t,s,L)$ be the minimal length of  LD $(\;s_L,\,\epsilon)$-codes of size~$t$.
If  $\epsilon=0$, then the number
 \beq{R0}
 R_L(s)\eq \varlimsup_{N\to\infty}\frac{\log_2 t_{0}(N,s,L)}{N}\,=\,
  \varlimsup_{t\to\infty}\frac{\log_2 t}{N_{0}(t,s,L)}
 \eeq
is called~\cite{dr83} the rate of LD $s_L$-codes.
\medskip

Observe~\cite{d14_pit} that  at fixed  $s\ge2$,
the number
\beq{limR_L}
R_{\infty}(s)\,\eq\,
\lim\limits_{L\to\infty}\,R_L(s), \qquad s=2,3,\dots,
\eeq
can be interpreted as the  {\em maximal rate} for two-stage group testing in the disjunctive search model
of any $d$, $d\le s$, defective elements  based on LD $s_L$-codes.
For the general two-stage group testing~\cite{d00_2}, the number $R_{\infty}(s)$ gives a lower bound on
the corresponding rate.

\textbf{Definition 4.}\quad Define the number
 \beq{R0+}
 C_L(s)\eq\varlimsup_{\epsilon\to0} \varlimsup_{N\to\infty}\frac{\log_2 t_{\epsilon}(N,s,L)}{N}\,=\,
 \varlimsup_{\epsilon\to0}  \varlimsup_{t\to\infty}\frac{\log_2 t}{N_{\epsilon}(t,s,L)}\,\ge\, R_L(s)
 \eeq
 called a {\em capacity} of almost disjunctive LD $s_L$-codes.
\medskip

The  definition~\req{R0+} implies  that if  the parameter $N$ is sufficiently large, then
for any fixed $\epsilon$, $\epsilon>0$, and any fixed  rate $R>0$,
 there exists an LD  $(\;s_L,\,\epsilon)$-code $X$ of length $N$
and  size $t=\lceil 2^{RN}\rceil$, i.e., $(N,R)$-code $X$,  if and only if the  rate~$R<C_L(s)$.
Obviously,  $C_L(s)\le 1/s$  and the first open problem is: "how to improve this evident upper bound?"


\textbf{Definition 5.}\quad
Let $R$, $R_L(s)\le R<C_L(s)$, be a fixed parameter. Taking into account the inequality~\req{B_eps}
from Definition~2, we
introduce the concept of  {\em error probability for almost disjunctive LD $s_L$-codes}:
\beq{e}
\epsilon_L(s,R,N)\eq\min\limits_{X\,:\,t=\left\lceil2^{RN}\right\rceil}\,\left\{\frac{\B_L(s,X)}{{t\choose s}}\right\},
\eeq
where the minimum is taken over all $(N,R)$-codes $X$, and  the function
\beq{E}
\E_L(s,R)\,\eq\,\varlimsup_{N\to\infty}\,\frac{-\log_2\epsilon_L(s,R,N)}{N},\quad
R_L(s)\le R<C_L(s),
\eeq
is said to be the  {\em exponent} of error probability for almost disjunctive LD $s_L$-codes.

Immediately from definitions~\req{R0}-\req{E}  it follows

\textbf{Proposition 1.}\quad {\em If one put the parameter $R=R_L(s)$, where the number $R_L(s)$  is
the rate of LD $s_L$-codes defined by~$\req{R0}$, then for the
exponent of error probability~$\req{E}$,  the  equality
\beq{ER0}
\E_L\left(s\,,\,R_L(s)\right)\,=\,s\,R_L(s), \quad s\ge1,\quad L\ge1
\eeq
holds. In other words, the value $R=R_L(s)$ is the unique root of the equation
\beq{ER}
\E_L\left(s\,,\,R\right)\,=\,s\,R, \quad s\ge1,\quad L\ge1.
\eeq
In addition,
 for any $s\ge1$ and $L\ge1$, the  monotonicity inequalities hold}:
\beq{RE-mon}
 R_L(s)\,\le\, R_{L+1}(s),\; C_L(s)\,\le\, C_{L+1}(s),\quad
 \E_L(s,R)\,\le\, \E_{L+1}(s,R).
\eeq


One can easily understand that Proposition 1 yields

\textbf{Proposition 2.}\quad Let there exist a lower bound $\underline{\E}_L(s,R)$
(upper bound $\overline{\E}_L(s,R)$) on the  exponent of error probability $\E_L(s,R)$
for almost disjunctive $s_L$-codes, i.e.,
$$
\underline{\E}_L(s,R)\le\,\E_L(s,R)\,\le\overline{\E}_L(s,R), \quad R_L(s)\le R<C_L(s).
$$
Let  $\underline{R}_L(s)$  ($\overline{R}_L(s)$) denote the unique root of the
equation
\beq{ER1}
 \underline{\E}_L(s,R)=s\,R,\quad (\overline{\E}_L\left(s\,,\,R\right)\,=\,s\,R), \quad s\ge1,\quad L\ge1.
\eeq
Then the number $\underline{R}_L(s)$  ($\overline{R}_L(s)$) is a lower (upper) bound
on the rate of LD $s_L$-codes, i.e. the inequality
$$
\underline{R}_L(s)\le R_L(s),\qquad (R_L(s)\le \overline{R}_L(s)),\quad s\ge1,\quad L\ge1.
$$
holds.

In Definitions~2-5 for the case $L=1$, we use the terminology which is similar to a terminology for the concept of weakly
 separating designs introduced in~\cite{m78}.  
Let $X$  be a code of length $N$ and size $t$ and let $\Omega_{\epsilon}(X,s,t)$ be a collection of
$s$-subsets of columns of the code $X$ such that its size $|\Omega_{\epsilon}(X,s,t)|\ge\,(1-\epsilon)\cdot\,{t\choose s}$.
 The  code $X$ is said~\cite{m78} to be  a {\em disjunctive  $(s,\epsilon)$-design}
 (or {\em weakly separating $s$-design}),
if there exists a collection $\Omega_{\epsilon}(X,s,t)$ such that the {\em disjunctive sums of any two $s$-subsets from
the collection $\Omega_{\epsilon}(X,s,t)$   are  different}. Weakly separating $s$-design can be considered~\cite{dr89}
(see, also~\cite{d03}) as an important example  of  information-theoretical model  for the  multiple-access channel~\cite{ck85}.
It was proved~\cite{m78} that the capacity of weakly separating $s$-designs  is equal to~$1/s$.
For the case $L\ge2$, the list-decoding weakly separating $s$-designs were suggested  in the paper~\cite{d81}, where it was established that
their capacity  is equal to~$1/s$ as well.

\section{Lower Bounds on $R_L(s)$, $C_L(s)$ and $\E_L(s,R)$}
\quad
The best known  upper and lower bounds on the rate  $R_L(s)$ of LD $s_L$-codes were
presented in~\cite{d14_pit} (see, also, preprint~\cite{d14}).
For the classical case $L=1$, these bounds have the form:
\beq{upR1-as}
R_1(s)\,\le\,\overline{R}_1(s)\,=\,\frac{2\log_2s}{s^2}\,(1+o(1)), \quad
s\to\infty,
\eeq
\beq{lowR1-as}
R_1(s)\ge\underline{R}_1(s)=\frac{4e^{-2}\log_2{s}}{s^{2}}(1+o(1))=
\frac{0,542\,\log_2{s}}{s^{2}}(1+o(1)),\qquad s\to\infty.
\eeq
If $s\ge1$, $L\ge2$, then our lower random coding bound on $R_L(s)$ was established~\cite{d14_pit} as

\textbf{Theorem 1.}~\cite{d14_pit}\quad
(Random coding bound~$\underline{R}^{(1)}_L(s)$).
{\bf1.}\quad  {\em The rate}
\beq{1}
R_L(s)\ge\underline{R}^{(1)}_L(s)\,\eq\,\frac{1}{s+L-1}\max\limits_{0<Q<1}\,A_L(s,Q)
=\frac{1}{s+L-1}\,A_L\left(s,Q^{(1)}_L(s)\right),
\eeq
\beq{AFromLDLowerBound}
 A_L(s,Q)\eq\log_2\frac{Q}{1-y}-sK(Q,1-y)-L\,K\left(Q,\frac{1-y}{1-y^s}\right),
\eeq
\beq{Kul}
K(a,b)\eq a\cdot \log_2 \frac{a}{b}+(1-a)\cdot \log_2 \frac{1-a}{1-b},\quad 0<a,b<1,
\eeq
{\em where 
parameter $y$, $1-Q\le y<1$, is defined as the unique root of the equation}
\beq{yFromLDLowerBound}
y=1-Q+Qy^s\left[1-\left(\frac{y-y^s}{1-y^s}\right)^L\right],
\qquad 1-Q\le y<1.
\eeq
{\bf2.}\quad {\em For fixed $L=2,3,\dots$ and $s\to\infty$, the asymptotic behavior
of the random coding bound  $\underline{R}^{(1)}_L(s)$ has the form}
\beq{ranL-4}
\underline{R}^{(1)}_L(s)=\frac{L}{s^2\log_2 e}(1+o(1))=\frac{L\,\ln2}{s^2}(1+o(1)).
\eeq
{\bf3.}\quad
{\em At fixed  $s=1,2,3,\dots$ and $L\to\infty$, 
 for the maximal rate $R_{\infty}(s)$ of two-stage group testing defined by~$\req{limR_L}$,
the lower bound
\beq{R1AsymptAsL}
R_{\infty}(s)\,\ge\,
\underline{R}^{(1)}_{\infty}(s)\,\eq\,\lim\limits_{L\to\infty}\,\underline{R}^{(1)}_{L}(s)=\log_2 \[ \frac{(s-1)^{s-1}}{s^s} + 1 \].
\eeq
holds. If $s\to\infty$, then}
$\underline{R}^{(1)}_{\infty}(s)=\frac{\log_2 e}{e\cdot s}(1+o(1))=\frac{0,5307}{s}(1+o(1))$.
\medskip

In the given paper, we suggest  a modification of the random coding method developed
in~\cite{d14_pit}
and obtain a lower bound on the capacity $C_L(s)$ along with a
lower bound on the exponent of error probability~$\E_L(s,R)$
for almost disjunctive $s_L$-codes. Let
 $$
[x]^+\eq\begin{cases}
x & \text{if $x\ge0$}, \cr
0 & \text{if $x<0$},\cr
\end{cases}
\quad\text{and}\quad
h(a)\eq -a\log_2a-(1-a)\log_2(1-a),\; 0<a<1,
$$
be the standard notations for the positive part function and the binary entropy function.
\medskip

\textbf{Theorem 2.}\quad
 (Random coding lower bounds~$\underline{C}(s)$ and
 $\,\underline{\E}_L(s,R)$).
{\em The following three claims hold.}
{\bf Claim 1.}\quad  {\em  The  capacity $C_L(s)$ and the exponent of error probability
$\E_L(s,R)$ for almost disjunctive LD $\;s_L$-codes satisfy inequalities
\beq{ranL0}
C_L(s)\,\ge\,\underline{C}(s)\,\eq\max\limits_{0<Q<1}\,C(s,Q)=C(s,Q(s)),\qquad s\ge1,\quad L\ge1,
\eeq
\beq{CQ}
C(s,Q)\eq\,h(Q)-\left[1-(1-Q)^s\right]\,h\left(\frac{Q}{1-(1-Q)^s}\right),\quad s\ge1,\quad 0<Q<1,
\eeq
and}
\beq{ranL1}
\E_L(s,R)\,\ge\,\underline{\E}_L(s,R)\,\eq\,\max\limits_{0<Q<1}\,E_L(s,R,Q),\qquad s\ge1,\quad L\ge1,
\eeq
\beq{EQ}
E_L(s,R,Q)\,\eq\,\min\limits_{Q\le q\le \min\{1, sQ\}}\;
\left\{\A(s,Q,q)+L\cdot[h(Q)-q\cdot h(Q/q)-R]^+\right\}.
\eeq
{\em where
the function $\A(s, Q, q)$,  $Q<q<\min\{1, sQ\}$,
is defined in  the  parametric  form:}
 \beq{Ay}
\A(s, Q, q) \eq (1-q) \log_2(1-q) + q \log_2 \[ \frac{Qy^s}{1-y} \] + sQ \log_2 \frac{1-y}{y} + sh(Q),
\eeq
\beq{ySmall}
q=Q\frac{1-y^s}{1-y},\qquad 0<y<1.
\eeq
{\bf Claim 2.}\quad {\em If $s\ge1$ is fixed, then  the random coding lower bound  $\underline{C}(s)>\frac{\ln2}{s}$
and at $s\to\infty$ the asymptotic behavior  of $\underline{C}(s)$ and the asymptotic behavior  of the optimal value
 $Q(s)$ in~$\req{ranL0}$ are:}
\beq{ranL-4}
\underline{C}(s)\,=\frac{\ln2}{s}(1+o(1)),\quad Q(s)\,=\frac{\ln2}{s}(1+o(1)).
\eeq
{\bf Claim 3.}\quad {\em For any $s\ge1$ and $L\ge1$, the lower bound
$\underline{E}_L(s, R)$ defined by~$\req{ranL1}$-$\req{ySmall}$
is a $\cup$-convex
function of the rate parameter $R>0$. If $\,0<R<\underline{C}(s)$, then $\underline{E}_L(s,R)>0$.
If $R\ge\underline{C}(s)$, then $\underline{E}_L(s,R)=0$. In
addition, there exist a number $R^{(cr)}_L(s)$, $0 \le R^{(cr)}_L(s) < \underline{C}(s)$, such
that
\beq{EBad}
\underline{E}_L(s, R) = (s + L - 1)\underline{R}_L^{(1)}(s) - LR, \quad\text{if}\quad 0 \le R \le R^{(cr)}_L(s),
\eeq
and
\beq{EGood}
\underline{E}_L(s, R) > (s + L - 1)\underline{R}_L^{(1)}(s) - LR, \quad \text{if} \quad R > R^{(cr)}_L(s),
\eeq
where the random coding bound $\underline{R}_L^{(1)}(s)$ is given
by the formulas~$\req{1}$-$\req{yFromLDLowerBound}$.}
\medskip

Table~1 gives some numerical values of the function
$$
\underline{R}_L(s)\,\eq\,\max\left\{\,\underline{R}_1(s)\,,\,\underline{R}^{(1)}_L(s)\,\right\},\quad
2\le s\le10, \quad 2\le L\le10,
$$
along with the corresponding  values $Q_L(s)$ of the optimal relative
weight $Q^{(1)}_L(s)$ in the right-hand side of~\req{1} if
$\underline{R}_L(s)=\underline{R}^{(1)}_L(s)$,
or we put $Q_L(s)\eq*\;$ if $\underline{R}_L(s)=\underline{R}_1(s)$,
where the values $\underline{R}_1(s)$ were calculated
in~\cite{d14_pit},
i.e,
$$
Q_L(s)\eq\begin{cases}
Q^{(1)}_L(s) & \text{if $\underline{R}_L(s)=\underline{R}^{(1)}_L(s)$\;for\; ($2\le s\le 6$, $L=2$)\;or\;($2\le s\le 10$, $3\le L\le10$) },\cr
* & \text{if $\underline{R}_L(s)=\underline{R}_1(s)$ \;for\;  ($7\le s\le 10$, $L=2$) }.\cr
\end{cases}
$$
The function $\underline{R}_L(s)$, $L\ge2$, $s\ge2$, can be considered as the best presently known lower bound
on the rate  $R_L(s)$, $L\ge2$, $s\ge2$, of LD $s_L$-codes.

\begin{table}[p]
\caption{}
\label{table1}
\begin{tabular*}{\textwidth}{@{\extracolsep{\fill}}cccccccccc}
\hline
$s_L$ & $2_2$ & $2_3$ & $2_4$ & $2_5$ & $2_6$ & $2_7$ & $2_8$ & $2_9$ & $2_{10}$ \cr
$Q_L(s)$ & $0.244$ & $0.233$ & $0.226$ & $0.221$ & $0.218$ & $0.215$ & $0.212$ & $0.211$ & $0.209$ \cr
$\underline{R}_L(s)$ & $0.2358$ & $0.2597$ & $0.2729$ & $0.2813$ & $0.2871$ & $0.2915$ & $0.2948$ & $0.2975$ & $0.2997$ \cr
$R^{(cr)}_L(s)$ & $0.3355$ & $0.3279$ & $0.3242$ & $0.3226$ & $0.3218$ & $0.3216$ & $0.3215$ & $0.3215$ & $0.3216$ \cr
\hline
$s_L$ & $3_2$ & $3_3$ & $3_4$ & $3_5$ & $3_6$ & $3_7$ & $3_8$ & $3_9$ & $3_{10}$ \cr
$Q_L(s)$ & $0.176$ & $0.167$ & $0.161$ & $0.156$ & $0.152$ & $0.149$ & $0.147$ & $0.145$ & $0.143$ \cr
$\underline{R}_L(s)$ & $0.1147$ & $0.1346$ & $0.1469$ & $0.1552$ & $0.1611$ & $0.1656$ & $0.1690$ & $0.1718$ & $0.1741$ \cr
$R^{(cr)}_L(s)$ & $0.2177$ & $0.2109$ & $0.2065$ & $0.2036$ & $0.2017$ & $0.2006$ & $0.1998$ & $0.1994$ & $0.1992$ \cr
\hline
$s_L$ & $4_2$ & $4_3$ & $4_4$ & $4_5$ & $4_6$ & $4_7$ & $4_8$ & $4_9$ & $4_{10}$ \cr
$Q_L(s)$ & $0.139$ & $0.133$ & $0.128$ & $0.123$ & $0.120$ & $0.117$ & $0.115$ & $0.113$ & $0.111$ \cr
$\underline{R}_L(s)$ & $0.0684$ & $0.0838$ & $0.0941$ & $0.1014$ & $0.1068$ & $0.1110$ & $0.1143$ & $0.1170$ & $0.1192$ \cr
$R^{(cr)}_L(s)$ & $0.1632$ & $0.1580$ & $0.1542$ & $0.1514$ & $0.1494$ & $0.1479$ & $0.1468$ & $0.1460$ & $0.1455$ \cr
\hline
$s_L$ & $5_2$ & $5_3$ & $5_4$ & $5_5$ & $5_6$ & $5_7$ & $5_8$ & $5_9$ & $5_{10}$ \cr
$Q_L(s)$ & $0.115$ & $0.110$ & $0.106$ & $0.103$ & $0.100$ & $0.098$ & $0.096$ & $0.094$ & $0.092$ \cr
$\underline{R}_L(s)$ & $0.0456$ & $0.0575$ & $0.0660$ & $0.0723$ & $0.0771$ & $0.0809$ & $0.0840$ & $0.0865$ & $0.0886$ \cr
$R^{(cr)}_L(s)$ & $0.1311$ & $0.1271$ & $0.1240$ & $0.1216$ & $0.1197$ & $0.1183$ & $0.1171$ & $0.1162$ & $0.1155$ \cr
\hline
$s_L$ & $6_2$ & $6_3$ & $6_4$ & $6_5$ & $6_6$ & $6_7$ & $6_8$ & $6_9$ & $6_{10}$ \cr
$Q_L(s)$ & $0.098$ & $0.095$ & $0.092$ & $0.089$ & $0.086$ & $0.084$ & $0.083$ & $0.081$ & $0.080$ \cr
$\underline{R}_L(s)$ & $0.0325$ & $0.0420$ & $0.0490$ & $0.0544$ & $0.0587$ & $0.0621$ & $0.0649$ & $0.0672$ & $0.0692$ \cr
$R^{(cr)}_L(s)$ & $0.1098$ & $0.1067$ & $0.1041$ & $0.1021$ & $0.1004$ & $0.0991$ & $0.0980$ & $0.0971$ & $0.0963$ \cr
\hline
$s_L$ & $7_2$ & $7_3$ & $7_4$ & $7_5$ & $7_6$ & $7_7$ & $7_8$ & $7_9$ & $7_{10}$ \cr
$Q_L(s)$ & $*$ & $0.083$ & $0.080$ & $0.078$ & $0.076$ & $0.074$ & $0.073$ & $0.072$ & $0.070$ \cr
$\underline{R}_L(s)$ & $0.0260$ & $0.0321$ & $0.0380$ & $0.0426$ & $0.0463$ & $0.0494$ & $0.0519$ & $0.0541$ & $0.0559$ \cr
$R^{(cr)}_L(s)$ & $0.0945$ & $0.0920$ & $0.0899$ & $0.0882$ & $0.0868$ & $0.0855$ & $0.0845$ & $0.0837$ & $0.0829$ \cr
\hline
$s_L$ & $8_2$ & $8_3$ & $8_4$ & $8_5$ & $8_6$ & $8_7$ & $8_8$ & $8_9$ & $8_{10}$ \cr
$Q_L(s)$ & $*$ & $0.074$ & $0.072$ & $0.070$ & $0.068$ & $0.067$ & $0.065$ & $0.064$ & $0.063$ \cr
$\underline{R}_L(s)$ & $0.0213$ & $0.0253$ & $0.0303$ & $0.0343$ & $0.0376$ & $0.0403$ & $0.0426$ & $0.0446$ & $0.0463$ \cr
$R^{(cr)}_L(s)$ & $0.0830$ & $0.0810$ & $0.0793$ & $0.0778$ & $0.0765$ & $0.0754$ & $0.0745$ & $0.0737$ & $0.0730$ \cr
\hline
$s_L$ & $9_2$ & $9_3$ & $9_4$ & $9_5$ & $9_6$ & $9_7$ & $9_8$ & $9_9$ & $9_{10}$ \cr
$Q_L(s)$ & $*$ & $0.067$ & $0.065$ & $0.063$ & $0.062$ & $0.061$ & $0.059$ & $0.058$ & $0.057$ \cr
$\underline{R}_L(s)$ & $0.0178$ & $0.0205$ & $0.0248$ & $0.0283$ & $0.0312$ & $0.0336$ & $0.0357$ & $0.0375$ & $0.0391$ \cr
$R^{(cr)}_L(s)$ & $0.0741$ & $0.0724$ & $0.0709$ & $0.0696$ & $0.0685$ & $0.0676$ & $0.0667$ & $0.0660$ & $0.0654$ \cr
\hline
$s_L$ & $10_2$ & $10_3$ & $10_4$ & $10_5$ & $10_6$ & $10_7$ & $10_8$ & $10_9$ & $10_{10}$ \cr
$Q_L(s)$ & $*$ & $0.061$ & $0.059$ & $0.058$ & $0.057$ & $0.056$ & $0.054$ & $0.054$ & $0.053$ \cr
$\underline{R}_L(s)$ & $0.0151$ & $0.0169$ & $0.0206$ & $0.0237$ & $0.0263$ & $0.0285$ & $0.0304$ & $0.0320$ & $0.0335$ \cr
$R^{(cr)}_L(s)$ & $0.0668$ & $0.0654$ & $0.0642$ & $0.0631$ & $0.0621$ & $0.0612$ & $0.0605$ & $0.0598$ & $0.0592$ \cr
\hline
$s$ & $2$ & $3$ & $4$ & $5$ & $6$ & $7$ & $8$ & $9$ & $10$ \cr
$\underline{C}(s)$ & $0.3832$ & $0.2455$ & $0.1810$ & $0.1434$ & $0.1188$ & $0.1014$ & $0.0884$ & $0.0784$ & $0.0704$ \cr
$Q(s)$ & $0.2864$ & $0.2028$ & $0.1569$ & $0.1280$ & $0.1080$ & $0.0935$ & $0.0824$ & $0.0736$ & $0.0666$ \cr
$R^{(cr)}_1(s)$ & $0.3510$ & $0.2284$ & $0.1705$ & $0.1364$ & $0.1137$ & $0.0976$ & $0.0855$ & $0.0761$ & $0.0685$ \cr
\hline
\end{tabular*}
\end{table}



\section{On Constructions of Almost Disjunctive Codes}
\quad
For $L=1$, constructions of LD  $\;s_1\,$-codes (i.e classical disjunctive (superimposed) $s$-codes)
 based on the shortened Reed-Solomon codes were developed in~\cite{d00_1}-\cite{d00_2}. The
papers~\cite{d00_1}-\cite{d00_2}  significantly  extend the  optimal and suboptimal constructions
of   superimposed $s$-codes suggested in~\cite{ks64} and contain the detailed tables with
parameters of the best known classical disjunctive (superimposed) $s$-codes.
In addition,  the table~3 from~\cite{d00_2} along with the similar table presented in~\cite{d00_3}
gives a range of  parameters $(t,N,s,\epsilon)$ corresponding to the best known
LD  $(\;s_1,\,\epsilon)$-codes based on MDS codes.
In the recent paper~\cite{br13}, it was proved that for the given parameters,
the following parametric asymptotic equations
\beq{RS-MDS}
t=q^{\left\lfloor\frac{q}{\log_2q}\right\rfloor},\;
N=q(q+1),\; \epsilon=\epsilon(q)\to0\;\mbox{if}\; s<q\cdot\ln2, \;q\;\mbox{-prime power},\;
 q\to\infty,
\eeq
hold. Note that if $s\to\infty$ and $q\to\infty$, then the asymptotic behavior of the rate for
LD  $(\;s_1,\,\epsilon)$-codes with parameters~\req{RS-MDS} is
$$
\frac{\log_2t}{N}=\frac{1}{q}(1+o(1))=\frac{\ln2}{s}(1+o(1))
$$
and coincides with the asymptotic behavior of the random coding bound $\underline{C}(s)$ defined by~\req{ranL-4}.

\medskip

\section{Proof of Theorem 2}
\quad
\textbf{Proof of claim 1.}\quad
 For an arbitrary code $X$, the number $\B_L(s,X)$  of   $s_L$-bad subsets of columns
in the code $X$ can be represented in the form:
\beq{s_L1}
\B_L(s,X)\eq\sum\limits_{\S\in[t],|\S|=s}\,\psi_L(X,\S),\qquad
\psi_L(X,\S)\,\eq\,
\begin{cases}
1 & \text{if the set $\x(\S)$ is $s_L$-bad in $X$},\cr
0 & \text{otherwise}.\cr
\end{cases}
\eeq
Let $Q$, $0<Q<1$, be a fixed parameter.
Introduce the constant-weight  ensemble $\{N,t,Q\}$ of binary $( N\times t)$-matrices $X$,
 where each column $\x(j)$, $j\in [t]$, of  $X$ 
 is taken with replacement
 from the
set containing  $N\choose w$ binary columns of a given weight~$w\eq\lfloor QN\rfloor$.
From~\req{s_L1} it follows that for the ensemble $\{N,\lfloor2^{RN}\rfloor,Q\}$, the expectation
$\overline{\B_L(s,X)}$ of the number  $\B_L(s,X)$ is
$$
\overline{\B_L(s,X)}\,=\,{t\choose s}\,\Pr\left\{\x(S)\; \text{is}\; s_L\text{-bad in}\; (N,R)\text{-code}\; X\right\}.
$$
Therefore, the expectation of the error probability for almost disjunctive LD $s_L$-codes is
\beq{B_L}
{\cal E}_L^{(N)}(s,R,Q)\eq\,{t\choose s}^{-1}\,\overline{\B_L(s,X)}=\,
\Pr\left\{\x(S)\; \text{is}\; s_L\text{-bad in}\; (N,R)\text{-code}\;
X\right\}.
\eeq
The evident {\em random coding upper bound} on the error probability~\req{e} for almost disjunctive LD $s_L$-codes
is formulated as the following inequality:
\beq{eQ}
\epsilon_L(s,R,N)\eq\min\limits_{X\,:\,t=\lfloor2^{RN}\rfloor}\,\left\{\frac{\B_L(s,X)}{{t\choose s}}\right\}
\,\le\,{\cal E}_L^{(N)}(s,R,Q),\quad 0<Q<1.
\eeq

The expectation
${\cal E}_L^{(N)}(s,R,Q)$ defined by~\req{B_L} can be represented in the form
\beq{B_L3}
{\cal E}_L^{(N)}(s,R,Q)\,=\,\sum\limits_{k=\lfloor QN\rfloor}^{\min\{N, s\lfloor QN\rfloor\}}\,
\Pr\left\{\x(S)\; \text{is}\; s_L\text{-bad in}\; X\,\left/\,\left|\bigvee_{i\in\S}\x(i)\right|=k\,\right.\right\}
\,\P^{(N)}(s,Q,k),
\eeq
where we  applied the total probability formula  and introduced the notation
\beq{k}
\P^{(N)}(s,Q,k)\,\eq\,\Pr\left\{\left|\bigvee_{i\in\S}\x(i)\right|=k\right\},\quad
\lfloor QN\rfloor\le k\le \min \{N, s\lfloor QN\rfloor\}.
\eeq
For the  ensemble $\{N,t,Q\}$ and any $k$,
$\lfloor QN\rfloor\le k\le \min \{N, s\lfloor QN\rfloor\}$,
the conditional
probability  of  event~\req{s_L} is
\beq{SL/k}
\Pr\left\{\bigvee_{i\in\S}\x(i)\succeq \bigvee_{j\in\L}\x(j)
\left/\,\left|\bigvee_{i\in\S}\x(i)\right|=k\,\right.\right\}=
\left[\frac{{k\choose \lfloor QN\rfloor }}{{N\choose \lfloor
QN\rfloor}}\right]^L.
\eeq
In addition, with the help of  the {\em type} (or {\em composition})  terminology:
$$
\{n(\a)\},\quad \a\eq(a_1,a_2,\dots,a_s)\in\{0,1\}^s,\quad 0\le n(\a)\le N,\quad
\sum\limits_{\a}n(\a)=N,
$$
 the probability of event~\req{k} in the ensemble $\{N,t,Q\}$
 can be written as follows:
\beq{k1}
\P^{(N)}(s,Q,k)\,= \,{N \choose \lfloor QN\rfloor }^{-s}\cdot  \,
\sum\limits_{\req{Qk}}\frac{N!}{\prod_{\a}n(\a)!},\quad
\lfloor QN\rfloor\le k\le \min\{N, s\lfloor QN\rfloor\},
\eeq
and in the right-hand side of~\req{k1}, the sum is taken over all types $\{n(\a)\}$ 
provided that
\beq{Qk}
n(\0)=N-k,\qquad \sum\limits_{\a:\,a_i=1}n(\a)=\lfloor QN\rfloor \quad \text{for any }i\in[s].
\eeq
Let the function
\beq{A}
\A(s,Q,q)\eq\,\lim\limits_{N\to\infty}\,\frac{-\log_2\,\P^{(N)}(s,Q,\lfloor qN\rfloor)}{N},\quad Q\le q\le \min\{1, sQ\},
\eeq
denotes the exponent of the logarithmic asymptotic behavior for the probability of event~\req{k} calculated
by~\req{k1}-\req{Qk}.

Further, the representation~\req{B_L3}, the conditional probability~\req{SL/k}
and the standard union bound
$$
\Pr\left\{\bigcup\limits_i\,C_i\,\left/C\right.\right\}\,\le\,\min\left\{1\,;\,\sum\limits_i\Pr\{C_i/C\}\right\}
$$
lead  to the upper bound
\beq{B_L4}
{\cal E}_L^{(N)}(s,R,Q)\,\le\,\sum\limits_{k=\lfloor QN\rfloor}^{\min\{N, s\lfloor QN\rfloor\}}\,\P^{(N)}(s,Q,k)
\,\min\,\left\{1\,;\,
{t-s\choose L}\,
\left[\frac{{k\choose \lfloor QN\rfloor}}{{N\choose \lfloor
QN\rfloor}}\right]^L\right\},
\eeq
where the code size $t\eq\lfloor2^{RN}\rfloor$.
Inequality~\req{B_L4} and the random coding bound~\req{eQ}
 imply that  the error probability exponent~\req{E} satisfies the inequality
\beq{ER}
\E_L(s,R)\,\ge\,\underline{\E}_L(s,R)\,\eq\,\max\limits_{0<Q<1}\,E_L(s,R,Q),
\eeq
\beq{EQ}
E_L(s,R,Q)\,\eq\,\min\limits_{Q\le q\le \min\{1, sQ\}}\;
\left\{\A(s,Q,q)+L\cdot[h(Q)-q\cdot h(Q/q)-R]^+\right\}.
\eeq

In {\bf Appendix} we prove

\textbf{Lemma 1.}\quad {\em  Let $\lfloor QN\rfloor\le k\le \min\{N, s\lfloor
QN\rfloor\}$. For the conditional probability in the right-hand
side of~$\req{B_L3}$, the lower bound
\beq{Lem4}
\Pr\left\{\x(S)\; \text{is}\; s_L\text{-bad in}\; X\,\left/\,\left|\bigvee_{i\in\S}\x(i)\right|=k\,\right.\right\}\geq
D(s, L)\times\min\,\left\{1\,;\,
{t-s\choose L}\,
\left[\frac{{k\choose \lfloor QN\rfloor}}{{N\choose \lfloor QN\rfloor}}\right]^L\right\},
\eeq
holds, where $D(s, L)$ is some constant.}

Lemma~1 establishes  the asymptotic accuracy of the upper bound
in~\req{B_L4}, i.e., there exists
$$
\lim\limits_{N\to\infty}\,\frac{-\log_2{\cal E}_L^{(N)}(s,R,Q)}{N}\,=\,
E_L(s,R,Q), \quad R>0.
$$
where the function $E_L(s,R,Q)$, $R>0$,  defined
by~\req{EQ} can be interpreted as the {\em exponent of  random coding bound on error
probability  for almost disjunctive LD $s_L$-codes} in the
ensemble~$\{N,\lfloor2^{RN}\rfloor,Q\}$ of  constant weight codes.

The  analytical properties of the function~$\req{A}$ are formulated below as Lemmas~2-4. 
They will be  proved in {\bf Appendix}.

\textbf{Lemma 2.}\quad {\em The function $\A(s, Q, q)$ of the parameter $q$, $Q<q<\min\{1, sQ\}$,
defined by~$\req{A}$  can be represented in the  parametric
form~$\req{Ay}$-$\req{ySmall}$}.
{\em In addition, the function $\A(s, Q, q)$ is $\cup$-convex, monotonically decreases  in the interval
$(Q, 1 - (1 - Q)^s)$, monotonically increases  in the interval $(1 - (1 - Q)^s, \min\{1, sQ\})$
and its unique minimal value which is equal to $0$ is attained at $q=1-(1-Q)^s$, i.e.,}
$$
\min\limits_{Q<q<\min\{1, sQ\}}\,\A(s, Q, q)\,=\,\A(s, Q,\,1-(1-Q)^s)=0,\quad 0<Q<1.
$$

\textbf{Lemma 3.}\quad {\em For any fixed  $Q$,  $0<Q<1$, the function $q\cdot h(Q/q)$, $Q<q<\min\{1, sQ\}$,
is an $\cap$-convex and monotonically increases.}

\textbf{Lemma 4.}\quad {\em For fixed $Q$, $0<Q<1$, the function
\beq{Lem3}
\A(s, Q, q) + L \cdot [h(Q) - q \cdot h(Q/q)], \quad Q < q < \min\{1, sQ\},
\eeq
is $\cup$-convex and its unique minimum is attained at some point $q = q^{(2)}_L(s, Q) > 1 - (1 - Q)^s$ and is equal to the function $A_L(s, Q)$,
defined by~$\req{AFromLDLowerBound}$-$\req{yFromLDLowerBound}$, i.e.,}
$$
\min \limits_{Q < q < \min\{1, sQ\}} \, \left\{ \A(s, Q, q) + L \cdot [h(Q) - q \cdot h(Q/q)] \right\} \, = \, A_L(s, Q).
$$

Claim~1  is an evident consequence of Lemma~2. $\quad \square$

Claim~3 is based on Lemmas~2-4.

\textbf{Proof of Claim 2.}\quad
First of all, let us rewrite the formula \req{CQ} in a more convenient form:
\beq{CQas}
C(s, Q) = (1 - Q - (1 - Q)^s) \log_2 \[ 1 - \frac{Q (1 - Q)^{s-1}}{1 - (1 - Q)^s)} \] - Q \log_2 \[ 1 - (1 - Q)^s \] - (1 - Q)^s \log_2 \[ 1 - Q \].
\eeq

For $Q_o(s, a) = \frac{a}{s} (1 + o(1)), s \to \infty$, the asymptotic behavior of \req{CQas} is the following:
\beq{CQopt}
C(s, Q_o(s, a)) = \frac{-a \log_2 \[ 1 - e^{-a} \]}{s} (1 + o(1)), \quad s \to \infty.
\eeq
Taking the derivative with respect to $a$ one can easily verify that for $a = \ln 2$ the maximum
$$
\max \limits_{a > 0} \left\{ -a \log_2 \[ 1 - e^{-a} \] \right\} = \ln 2
$$
is attained. Thus,
\beq{Cge}
\underline{C}(s) \ge \frac{\ln 2}{s} (1 + o(1)), s \to \infty.
\eeq
To complete the proof we need to achieve the opposite asymptotic inequality.

Let $0 < Q(s) < 1,$ $s = 2, 3, ...,$ be an arbitrary sequence, such that
$$
\max \limits_{0 < Q < 1} C(s, Q) = C(s, Q(s)) = \underline{C}(s).
$$

Suggest that $Q(s) > b$, for some fixed $b > 0$. Then, one can obtain from \req{CQas} the inequality $C(s, Q(s)) \le (1-b)^s O(1), s \to \infty$, and there is a contradiction with \req{Cge}. Hence, without loss of generality, $Q(s) \to 0$, as $s \to \infty$.

Suggest that $0 < Q = f(s) / s < 1$ and $\lim_{s \to \infty} f(s) = \infty$. The assumption yields
$$
\lim_{s \to \infty} (1 - Q)^s \le \lim_{s \to \infty} e^{-f(s)} = 0.
$$
Using the previous property and the expansion of a logarithm one can derive from \req{CQas} the asymptotic inequality $C(s, Q(s)) \le Q (1 - Q)^s O(1), s \to \infty$. Therefore, the equality
$$
\lim_{s \to \infty} sQ (1 - Q)^s = 0
$$
establishes a contradiction with \req{Cge}. Thus, without loss of generality, $sQ(s) \to a$, as $s \to \infty$, where the condition $0 \le a < \infty$ holds.

Similarly, the assumption $a = 0$ leads to the asymptotic inequality $C(s, Q(s)) \le Q O(1)$, where is a contradiction with \req{Cge}.

Thus, the asymptotics \req{ranL-4} holds. Claim 2 is proved. $\quad \square$.

\textbf{Proof of Claim 3.}\quad
Note that the $\cup$-convexity of $E_L(s, R, Q)$ for arbitrary $0 < Q < 1$ implies the $\cup$-convexity of $\underline{E}_L(s, R)$. Let us prove the $\cup$-convexity of $E_L(s, R, Q)$.

Fix arbitrary $0 < Q < 1$. For a fixed $R > 0$, it follows from Lemmas~2-4 that the minimum in \req{EQ} is attained at some point $q \in [\, q^{(0)}(s, Q) = 1 - (1 - Q)^s, q^{(2)}_L(s, Q) \,]$. Denote $\BB(R, Q, q) = h(Q) - qh(Q / q) - R$. If there exists a solution $q \in (0, 1)$ of the equation $\BB(R, Q, q) = 0$, we will denote it as $q^{(1)}(R, Q)$. It's clear that the minimum in \req{EQ} is attained at the point $q = q^{(min)}_L(s, R, Q)$, defined as
$$
q^{(min)}_L(s, R, Q) =
\begin{cases}
q^{(2)}_L(s, Q), \quad &\text{if } \BB(R, Q, q^{(2)}) >= 0,\\
q^{(1)}(R, Q), \quad &\text{if } \BB(R, Q, q^{(0)}) > 0 \text{ and } \BB(R, Q, q^{(2)}) < 0,\\
q^{(0)}(s, Q), \quad &\text{if } \BB(R, Q, q^{(0)}) <= 0.
\end{cases}
$$
Correspondingly, the substitution of $q^{(min)}$ into the expression \req{EQ} gives
\beq{EQmin}
E_L(s, R, Q) =
\begin{cases}
A_L(s, Q) - LR, \quad &\text{for } 0 \le R \le R_L^{(cr)}(s, Q),\\
\A(s, Q, q^{(1)}), \quad &\text{for } R_L^{(cr)}(s, Q) \le R \le C(s, Q),\\
0, \quad &\text{for } C(s, Q) \le R,
\end{cases}
\eeq
where $A_L(s, Q)$ is defined by \req{AFromLDLowerBound}-\req{yFromLDLowerBound}, $\A(s, Q, q)$ characterized by \req{Ay}-\req{ySmall}, $C(s, Q)$ determined by \req{CQ} and
\beq{RCriticalQ}
R_L^{(cr)}(s, Q) = h(Q) - q^{(2)} h(Q / q^{(2)}).
\eeq

Note that $q^{(1)}(R, Q)$ is the implicit function of the parameter $R$ defined by the equation $\BB(R, Q, q) = 0$. Hence, one can calculate the following derivative in the domain of the function $q^{(1)}(R, Q)$:
\beq{q1Derivative}
\( q^{(1)}(R, Q) \)^{\prime}_R = 1 \div \log_2 \frac{q - Q}{q}.
\eeq
Therefore, the use of \req{EQmin} and \req{q1Derivative} allows to compute the derivative of $E_L(s, R, Q)$ with respect to $R$:
$$
\( E_L(s, R, Q) \)^{\prime}_R =
\begin{cases}
-L, \quad &\text{for } 0 \le R \le R_L^{(cr)}(s, Q),\\
\log_2 \frac{Qy^s}{1 - Q - y + Qy^s} \div \log_2 \frac{q - Q}{q}, \quad &\text{for } R_L^{(cr)}(s, Q) \le R \le C(s, Q),\\
0, \quad &\text{for } C(s, Q) \le R,
\end{cases}
$$
where in the second line $q$ denotes $q^{(1)}(R, Q)$ and $y$ is defined by \req{ySmall}. One can easily verify that the expression in the second line is nondecreasing function of the parameter $R$, moreover it equals $-L$ at $R = R_L^{(cr)}(s, Q)$ and $0$ at $R = C(s, Q)$. Thus, the derivative of $E_L(s, R, Q)$ with respect to $R$ exists, is continuous and nondecreasing function, i.e. $E_L(s, R, Q)$ is $\cup$-convex.

In the case $R = 0$, for any $0 < Q < 1$, it is clear that $h(Q) - qh(Q / q) \ge 0$, therefore the case $R = 0$ satisfies \req{EBad}.

In the case $R = \underline{C}(s)$, it is clear that $\underline{E}_L(s, R) = 0$, therefore the case $R = \underline{C}(s)$ satisfies \req{EGood}.

Thus, due to the $\cup$-convexity of $\underline{E}_L(s, R)$, there exists $R^{(cr)}_L(s)$, such that \req{EBad} holds for $0 \le R \le R^{(cr)}_L(s)$ and \req{EGood} holds for $R > R^{(cr)}_L(s)$.

Claim~3 is proved. $\quad \square$.



\section{Appendix: Proofs of Lemmas 1-4}
\quad
\textbf{Proof of Lemma 1.}\quad
Denote $p=\left[\frac{{k\choose \lfloor QN\rfloor}}{{N\choose \lfloor QN\rfloor}}\right]$.
Let $A_i$ be an event, that $\x(S)$ covers $L$ fixed columns, $1\leq i\leq {t-s\choose L}$, then $\Pr(A_i)=p^L$.

$\Pr\left\{\x(S)\; \text{is}\; s_L\text{-bad in}\; X\,\left/\,\left|\bigvee_{i\in\S}\x(i)\right|=k\,\right.\right\}\geq
\sum\limits_{i=1}^{t-s\choose L} \Pr(A_i) - \sum\limits_{1\leq i < j\leq {t-s\choose L}} \Pr(A_iA_j)
$


$\sum\limits_{1\leq i < j\leq {t-s\choose L}} \Pr(A_iA_j)=\frac{{t-s\choose L}}{2}\sum\limits_{i=2}^{{t-s\choose L}}\Pr(A_1A_i)=\\
\frac{{t-s\choose L}}{2}\sum\limits_{l=0}^{L-1}{L \choose l}{t-s-L \choose L - l}\Pr(A_1A_j|\;\text{the cardinality of intersection } A_1A_j \text{ is equal to } l)=\\
\frac{{t-s\choose L}}{2}\sum\limits_{l=0}^{L-1}{L \choose l}{t-s-L \choose L - l} p^{2L-l}<
{t-s\choose L}p^L\sum\limits_{l=0}^{L-1}{L \choose L-l}(tp)^{L-l}<
{t-s\choose L}p^L((1+tp)^L-1).
$

Let $t_0$ be a root of the equation $(1+tp)^L-1=0.5$, i.e. $t_0=\frac{(1.5)^{\frac{1}{L}}-1}{p}$.

If $t<t_0$, then $(1+pt)^L<1.5$ and $$\Pr\left\{\x(S)\; \text{is}\; s_L\text{-bad in}\; X\,\left/\,\left|\bigvee_{i\in\S}\x(i)\right|=k\,\right.\right\}\geq
\frac{1}{2}{t-s\choose L}p^L.
$$
If $t>t_0>s+L$ then

$$\Pr\left\{\x(S)\; \text{is}\; s_L\text{-bad in}\; X\,\left/\,\left|\bigvee_{i\in\S}\x(i)\right|=k\,\right.\right\}\geq
\frac{1}{2}{t_0-s\choose L}p^L\geq\frac{1}{2}
\left(\frac{t_0p}{s+L+1}\right)^L=D_1(s, L).$$

If $t_0\leq s+L$, then $p\geq \frac{1.5^{\frac{1}{L}}-1}{s+L}$ and
$$\Pr\left\{\x(S)\; \text{is}\; s_L\text{-bad in}\; X\,\left/\,\left|\bigvee_{i\in\S}\x(i)\right|=k\,\right.\right\}\geq
D_2(s, L),$$ where $D_2(s, L)$ is some constant.

Setting
$D(s, L)=\min(D_1(s, L), D_2(s, L), 0.5)$ we obtain the inequality $\req{Lem4}$.

Lemma 1 is proved. $\quad \square$

\textbf{Proof of Lemma 2.}\quad
Let $s \ge 2$, $0 < Q < 1$, $Q < q < \min \{ 1, sQ \}$ be fixed parameters.
Let $k = \lfloor qN \rfloor$ and $N \to \infty$. For every type $\{n(\a)\}$ we will consider corresponding distribution $\tau: \tau(\a) = \frac{n(\a)}{N}, \quad \forall~\a \in \{0, 1\}^s$.

Applying the Stirling approximation, we obtain the following logarithmic asymptotic behavior of the summand in the sum \req{k1}:
$$
-\log_2 \frac{N!}{\prod_{\a}n(\a)!} {N \choose \lfloor QN \rfloor}^{-s} = N F(\tau, Q, q)(1 + o(1)), \quad \text{where}
$$
\beq{FNeedToMin}
F(\tau, Q, q) = \sum_{\a}\tau(\a) \log_2 \tau(\a) + sH(Q).
\eeq
Thus, one can reduce the calculation of $\A(s, Q, q)$ to the search of the following minimum:
\beq{FProblem}
\A(s, Q, q) = \min_{\tau \in \req{FRegion}: \req{FRestrictions}} F(\tau, Q, q),
\eeq
\beq{FRegion}
\left\{ \tau:~\forall~\a \quad 0 < \tau(\a) < 1 \right\},
\eeq
\beq{FRestrictions}
\sum_{\a} \tau(\a) = 1, \qquad \tau(\0) = 1 - q, \qquad \sum_{\a: a_i = 1} \tau(\a) = Q \quad \forall~i \in [s],
\eeq
where the restrictions \req{FRestrictions} are induced by the definition of type and the properties \req{Qk}.

To find the minimum, we use the standard Lagrange multipliers method. The Lagrangian is equal to
\begin{multline*}
\Lambda \eq \sum_{\tau(\a)} \tau(\a) \log_2 \tau(\a) + sh(Q) + \lambda_0 \(\tau(\0) + q - 1\) +\\
+ \sum_{i = 1}^s \lambda_i \(\sum_{\a: a_i = 1} \tau(\a) - Q\) + \lambda_{s+1} \(\sum_{\a} \tau(\a) - 1\).
\end{multline*}
Therefore, the necessary conditions for the extremal distribution are
\beq{FNecConditions}
\begin{cases}
\frac{\partial \Lambda}{\partial \tau(\0)} = \log_2 \tau(\0) + \log_2 e + \lambda_0 + \lambda_{s+1} = 0,\\
\frac{\partial \Lambda}{\partial \tau(\a)} = \log_2 \tau(\a) + \log_2 e + \lambda_{s+1} + \sum_{i = 1}^s a_i \lambda_i = 0 \quad \text{for any } \a \neq \0.
\end{cases}
\eeq

The matrix of second derivatives of the Lagrangian is obvious to be diagonal. Thus, this matrix is positive definite in the region \req{FRegion} and the function $F(\tau, Q)$ defined by \req{FNeedToMin} is strictly $\cup$-convex in the region \req{FRegion}. The Karush-Kuhn-Tacker theorem (see, for example, \cite{opu}) states that each solution $\tau \in \req{FRegion}$ satisfying system \req{FNecConditions} and constraints \req{FRestrictions} gives a local minimum of $F(\tau, Q)$. Thus, if there exists a solution of the system \req{FNecConditions} and \req{FRestrictions} in the region \req{FRegion}, then it is unique and gives a minimum in the minimization problem \req{FProblem} - \req{FRestrictions}.

Note that the symmetry of problem yields the equality $v \eq \lambda_1 = \lambda_2 = ... \lambda_s$. Let $u \eq \log_2 e + \lambda_{s+1}$ and $w \eq \lambda_0$. One can rewrite \req{FRestrictions} and \req{FNecConditions} as follows:
\beq{FFinalSystem}
\begin{cases}
\text{1) } \log_2 \tau(\a) + u + v \sum_{i = 1}^s a_i = 0 \quad \text{for any } \a \neq \0,\\
\text{2) } \log_2 \tau(\0) + u + w = 0,\\
\text{3) } \tau(\0) = 1 - q,\\
\text{4) } \sum_{\a} \tau(\a) = 1,\\
\text{5) } \sum_{\a: a_i = 1} \tau(\a) = Q \quad \text{for any } i \in [s].
\end{cases}
\eeq

Let $y \eq \frac{1}{1+2^{-v}}$ be a change of the variable $v$.
The first equation of the system \req{FFinalSystem} means that
\beq{FTauANotLast}
\text{for every } \a \neq \0 \quad \tau(\a) = \frac{1}{2^u y^s} (1-y)^{\sum a_j} y^{s - \sum a_j}.
\eeq
The substitution of \req{FTauANotLast} into the equation 5) of the previous system allows us to obtain
$$
\sum_{\a : a_i = 1} \frac{1}{2^u y^s} (1-y)^{\sum a_j} y^{s - \sum a_j} = \frac{1-y}{2^u y^s},
$$
and therefore, the solution $u$ is determined by the equality
\beq{FVarU}
u = \log_2 \frac{1 - y}{Qy^s}.
\eeq
Substituting \req{FTauANotLast}, \req{FVarU} and the third equation of \req{FFinalSystem} into the equation 4) of the system \req{FFinalSystem} we achieve
$$
q = \sum_{\a \neq 0} \tau(\a) = \frac{Q(1 - y^s)}{1 - y},
$$
i.e. the equation \req{ySmall}. Thus, the conditions \req{FRestrictions} and \req{FNecConditions} have the unique solution $\tau$ in the region \req{FRegion}:
\beq{FTauSolution}
\tau(\0) = 1 - q, \qquad \tau(\a) = \frac{Q}{1 - y} (1-y)^{\sum a_j} y^{s - \sum a_j} \quad \text{for any } \a \neq \0,
\eeq
where the parameters $q$ and $y$ are related by the equation \req{ySmall}.
To get the exact formula \req{Ay}, the substitution of \req{FTauSolution} into \req{FNeedToMin} is sufficient.


 Let us prove the properties of the function \req{Ay}.
Note that the function $q(y) = Q \frac{1 - y^s}{1 - y}$ \req{ySmall} monotonically increases in the interval $y \in (0, 1)$ and correspondingly takes the values $Q$ and $sQ$ at the ends of the interval. That is why one can consider the function \req{Ay} as the function $\T(s, Q, y) = \A(s, Q, q(y))$ of the parameter $y$ in the interval $y \in (0, y_1)$, where $q(y_1) = \min\{1, sQ\}$. The derivative of the function $\T(s, Q, y)$ equals
\beq{dLem1}
\T^{\prime}(s, Q, y) = q^{\prime}(y) \log_2 \frac{Qy^s}{1 - Q - y + Qy^s}.
\eeq
Thus, $\T(s, Q, y)$ decreases in the interval $y \in (0, 1 - Q)$,
increases in the interval $y \in (1 - Q, y_1)$, is $\cup$-convex,
attains the minimal value $0$ at  $y_0 = 1 - Q$ and $q(y_0) = 1 - (1 - Q)^s$.

Lemma 2 is proved. $\quad \square$

\textbf{Proof of Lemma 3.}\quad
Let $0 < Q < 1$ be a fixed value. The derivative of the function $f(Q, q) = q \cdot h(Q / q),~Q < q < 1$ equals
\beq{dLem2}
f^{\prime}_q(Q, q) = - \log_2 \frac{q - Q}{q}.
\eeq
Hence, the function $f(Q, q)$ increases in the interval $q \in (Q, 1)$, is $\cap$-convex and, for any half-interval $q \in (Q, a], Q < a < 1$, attains its unique maximal value at the point $q = a$.

Lemma 3 is proved. $\quad \square$

\textbf{Proof of Lemma 4.}\quad
Let $0 < Q < 1$ be a fixed value. Due to the properties of \req{ySmall}, one can consider the function \req{Lem3} as the function
$$
\F(s, L, Q, y) = \A(s, Q, q(y)) + L [h(Q) - q(y) \cdot h(Q / q(y))]
$$
of the parameter $0 < y < y_1$, where $q(y_1) = \min \{1, sQ\}$. Using \req{dLem1} and \req{dLem2} one can calculate the derivative of $\F(s, L, Q, y)$:
$$
\F^{\prime}(s, L, Q, y) = \T^{\prime}(s, Q, y) - L q^{\prime}(y) f^{\prime}_q(Q, y) = q^{\prime}(y) \cdot \log_2 \[ \frac{Qy^s}{1 - Q - y + Qy^s} \( \frac{y - y^s}{1 - y^s} \)^L \].
$$
Thus, the equality $\F^{\prime}(s, L, Q, y) = 0$ holds if and only if
$$
y = 1 - Q + Qy^s \[ 1 - \( \frac{y - y^s}{1 - y^s} \)^L \],
$$
i.e. the relation \req{yFromLDLowerBound} is true. The function \req{Lem3} is clear to be $\cap$-convex and to attain the minimum at the point $q = q(y_2)$, where $y_2$ is the solution of the equation \req{yFromLDLowerBound}.

Note that the following equality holds:
$$
1 - q(y_2) = 1 - \frac{Q(1 - y_2^s)}{1 - y_2} = \frac{Qy_2^s}{1 - y_2} \( \frac{y_2 - y_2^s}{1 - y_2^s} \)^L.
$$
Thus
\begin{multline*}
\F(s, L, Q, y_2) = \( 1 - Q \frac{1 - y_2^s}{1 - y_2} \) \log_2 \[ \frac{Qy_2^s}{1 - y_2} \( \frac{y_2 - y_2^s}{1 - y_2^s} \)^L \] + Q \frac{1 - y_2^s}{1 - y_2} \log_2 \frac{Qy_2^s}{1 - y_2} +\\
sQ \log_2 \frac{1 - y_2}{y_2} + sh(Q) + Lh(Q) + LQ \log_2 \frac{1 - y_2}{1 - y_2^s} + LQ \frac{y_2 - y_2^s}{1 - y_2} \log_2 \frac{y_2 - y_2^s}{1 - y_2^s}.
\end{multline*}
The simplifying of the previous expression yields
$$
\min \limits_{0 < y < y_1} \F(s, L, Q, y) = A_L(s, Q),
$$
where the function $A_L(s, Q)$ is defined by \req{AFromLDLowerBound}-\req{yFromLDLowerBound}.

Lemma 4 is proved. $\quad \square$


\newpage
\begin{thebibliography}{99}

\bibitem{f65}
\textit{Fano R.}
Transmission of Information. A Statistical Theory of Communications.
Wiley,  New York - London,  1961.

\bibitem{g68}
\textit{Gallager R.G.}
Information Theory and Reliable
Communication. J.~Wiley, New York, 1968.

\bibitem{ck85}
\textit{Csiszar I.},  \textit{Korner J.}
Information Theory. Coding Theorems for Discrete Memoryless Systems.
Akademiai Kiado, Budapest, 1981.



\bibitem{ks64}
\textit{Kautz W.H.}, \textit{Singleton R.C.}  Nonrandom Binary Superimposed
Codes//  IEEE Trans.  Inform. Theory. 1964. V.~10. n~4. P.~363-377.

\bibitem{dr82}
\textit{D'yachkov A.G.},  \textit{Rykov V.V.}
Bounds on the Length of Disjunctive Codes //
Problems of Information Transmission.
1982. V.~18. n~3. P.~166-171.

\bibitem{e82}
\textit{Erdos P.},  \textit{Frankl P.}, \textit{Furedi Z.}
Families of Finite Sets in Which No
 Set Is Covered by the Union of $2$ Others //
J.  Combin. Theory. Ser.~A. 1982. V.~33. P.~158-166.

\bibitem{dr83}
\textit{D'yachkov A.G.},  \textit{Rykov V.V.}
A Survey of Superimposed Code Theory  //
Problems of Control and Inform. Theory. 1983. V.~12.  n~4. P.~229-242.

\bibitem{drr89}
\textit{D'yachkov A.G.},  \textit{Rykov V.V.}, \textit{Rashad A.M.}
Superimposed Distance Codes //
Problems of Control and Inform. Theory. 1989. V.~18. n~4. P.~237-250.

\bibitem{d00_1}
\textit{D'yachkov A.G.}, \textit{Macula A.J.}, \textit{Rykov V.V.}
New Constructions of Superimposed Codes //
IEEE Trans.  Inform. Theory. 2000. V.~46. n~1. P.~284-290.

\bibitem{d00_2}
\textit{D'yachkov A.G.}, \textit{Macula A.J.}, \textit{Rykov V.V.}
New Applications and Results of Superimposed Code Theory Arising
from the Potentialities of Molecular Biology // In the
book "Numbers, Information and Complexity". P.~265-282,
Kluwer Academic Publishers, 2000.

\bibitem{d00_3}
\textit{D'yachkov A.G.}, \textit{Vilenkin P.A.}, \textit{Macula A.J.}, \textit{Torney D.C.}, \textit{Yekhanin S.M.}
New Results in the Theory of Superimposed Codes  //
Proc. Seventh Int. Workshop on Algebraic and Combinatorial Coding Theory. Bansko, Bulgaria. 2000. pp.~126-136.

\bibitem{d02}
\textit{D'yachkov A.},  \textit{Vilenkin P.}, \textit{Macula A.}, \textit{Torney D.}
Families of Finite Sets in Which No Intersection
of $\l$ Sets Is Covered by the Union of $s$ Others //
J.  Combin. Theory. Ser.~A. 2002. V.~99.
pp.~195-218.


\bibitem{d03}
\textit{D'yachkov A.G.}
Lectures on Designing
        Screening Experiments // Lecture Note Series~10, Feb.~2003,
        Combinatorial and Computational Mathematics Center,
        Pohang  University of Science and Technology (POSTECH),
        Korea Republic, (survey, 112 pages).


\bibitem{d14_pit}
\textit{D'yachkov A.G.}, \textit{Vorobyev I.V.}, \textit{Polyanskii N.A.},  \textit{Shchukin V.Yu.}
Bounds on the Rate of Disjunctive Codes // Problems of Information Transmission, vol. 50, no. 1, pp. 27-56, 2014.

\bibitem{d14}
\textit{D'yachkov A.G.}, \textit{Vorobyev I.V.}, \textit{Polyanskii N.A.},  \textit{Shchukin V.Yu.}
Bounds on the Rate of Superimposed Codes// arXiv: 1401.0050 [cs.IT].

\bibitem{m78}
\textit{Malyutov M.B.}
The Separating Property of Random Matrices // Mathematical Notes. 1978. V.23. n~1. P.~84-91.




\bibitem{dr89}
\textit{D'yachkov A.G.}, \textit{Rashad A.M.}
Universal Decoding for Random Design of Screening Experiments //
 Microelectronics and Reliability. 1989. V.~29. n~6. P.~965-971.

\bibitem{d81}
\textit{D'yachkov A.G.}
 Error probability bounds for the symmetrical model of the design of screening experiments//
Problems of Information Transmission. 1981. V.~17 n.~4. pp.~245-263.



\bibitem{br13}
\textit{Bassalygo L.A.}, \textit{Rykov V.V.}
Multiple-access hyperchannel // Problems of Information Transmission, 2013. vol. 49. no. 4, pp. 299-307.

\bibitem{opu}
\textit{Galeev E.M.}, \textit{Tikhomirov V.M.}
Optimization: theory, examples, problems.
Editorial URSS, Moscow, 2000. (in Russian)













\end{thebibliography}
\end{document}